\documentclass[prl,aps,twocolumn,groupedaddress,showpacs,floatfix]{revtex4}
\usepackage{amsmath,amssymb,multirow,epsfig,bm}
\usepackage{psfrag}
\usepackage{subfigure}

\newcommand{\beq}{\begin{equation}}
\newcommand{\eeq}{\end{equation}}
\newcommand{\bea}{\begin{eqnarray}}
\newcommand{\eea}{\end{eqnarray}}

\begin{document}
\title {Critical Behavior in Graphene with Coulomb Interactions}

\author{Jianhui Wang$^1$, H.A. Fertig$^1$, and Ganpathy Murthy$^2$}
\affiliation{$^1$Department of Physics, Indiana University, Bloomington, IN 47405}
\affiliation{$^2$Department of Physics and Astronomy, University of Kentucky, Lexington, KY 40506-0055}

\begin{abstract}
We demonstrate that in the presence of Coulomb interactions,
electrons in graphene behave like a critical system, supporting
power law correlations with interaction-dependent exponents.
An asymptotic analysis shows that the origin of this behavior
lies in particle-hole scattering, for which
the Coulomb interaction induces anomalously close approaches.
With increasing interaction strength the relevant power
law changes from real to complex, leading to an unusual
instability characterized by a complex-valued susceptibility
in the thermodynamic limit.
Measurable quantities, as well as the connection to
classical two dimensional systems, are discussed.
\end{abstract}

\date{\today}

\maketitle

Graphene continues to fascinate physicists with its many unique
properties \cite{GeimNovoselov,neto109}.  The low energy physics of
electrons in graphene may be described by a Dirac
Hamiltonian, analogous to the theory of massless neutrinos,
with the two components of the electron wavefunctions
representing amplitudes on the two sublattices
that make up the graphene lattice.
One class of problems
yet to be understood in a fuller description
involves the effect
of the Coulomb interaction.  Its strength may be
characterized by an effective fine structure constant
$\beta=e^2/\epsilon \hbar v_F$, where $v_F$ is the electron
speed near the Dirac point and $\epsilon$ is a dielectric
constant due to a substrate.
For suspended
graphene, $\beta$ is estimated to be of order 2.  Naively, then,
Coulomb interactions are relatively strong in graphene, so
that one may expect to see its effects in a clean enough sample.
This is the subject of our study.

In the presence of Coulomb interactions, the Hamiltonian
of this system, remarkably, has no natural length scale:
the $1/r$ interaction has
precisely the same operator dimension
as the (Dirac) kinetic energy.  This suggests
that the system behaves as if it is at a critical
point, even though no parameters need be tuned to attain
this situation.    In what follows, we
demonstrate that a dramatic effect of Coulomb interactions
is that they induce power law correlations,
a hallmark property of
critical systems.  The underlying
cause of this originates in short distance physics
-- an anomalously large probability for close approaches
of particle-hole pairs -- but consequences are
manifested at long distances because of the absence of
a length scale in the Hamiltonian.  The power law
correlations in this system are reminiscent of the
behavior of a variety of {\it classical} two-dimensional
systems \cite{kosterlitz73}, and like those, when
the coupling constant is sufficiently large we find
indications of an unusual phase transition, 
characterized in the thermodynamic limit by
a susceptibility that goes from real to complex
rather than diverging. We speculate that resulting
state may represent a precursor to
the formation of a (gapped) exciton
condensate \cite{khveshchenko,gorbar2002}.  

The many-body physics we describe below is present in
a simpler form for non-interacting Dirac electrons
scattering from a Coulomb potential
$V(r)=-Ze^2/r$ \cite{shytov2007,pereira166802,biswas205122,terekhov076803,pereira085101,kotov075433}.
For small $r$, the wavefunctions vanish as $\psi_m(r) \sim
r^{\sqrt{(m+1/2)^2-Z^2\beta^2}-1/2}$.  In contrast,
for impurity potentials which do not
diverge so strongly at $r=0$, the power law
is fixed by the ``centrifugal barrier'' associated
with a given angular momentum channel $m$, and does not
depend on the potential itself.  The
fact that the exponent becomes a {\it continuously}
varying function of $\beta$ is unusual: the $1/r$ attraction
allows an anomalous
penetration of
the centrifugal barrier.  Moreover, for $Z\beta \ge m$,
the exponent becomes complex, and
one must
introduce a short distance cutoff to obtain
sensible wavefunctions.  This
``Coulomb implosion'' phenomenon is accompanied
by the appearance of a screening cloud $\rho(r)
\sim 1/r^2$ around the impurity which is not present for smaller
$\beta$.

General particle-hole channel propagators support
analogous behavior, generating
power law behavior in certain correlation functions.
We show in particular that the
sublattice-antisymmetric susceptibility does this,
so that critical-like behavior
is manifested in the response to potentials
which are not symmetric for the two sublattices.
Moreover we
find a divergence
when $\beta$ exceeds
$\beta_c$, when the centrifugal barrier
in particle-hole scattering is overcome and
the power changes from real to complex,
a many-body manifestation of Coulomb implosion.


{\it Analysis in Momentum Representation} -- To motivate our
approach, we begin by analyzing the
noninteracting problem in the presence of a Coulomb impurity as
a scattering problem, using a standard momentum representation. The Hamiltonian is
$
H_0 \psi=[\varepsilon-V(\rho)]\psi,
$
where $H_0=\hbar v_F \vec{\sigma}\cdot\hat{\vec{p}}$ is the kinetic
energy for one of the two valleys, with $\vec{\sigma}$ the Pauli
matrices acting on the space of the two sublattices, $\hat{\vec{p}}$
the momentum operator, and $V(\rho)$ the Coulomb impurity potential.  This is a low-energy
effective Hamiltonian valid at distances large compared to the lattice
scale; we ignore the small separation between different
sublattice points of the same lattice site. In clean noninteracting
graphene there is negligible intervalley scattering. This remains
approximately true here as well because the
valleys are separated by a large momentum, and the Coulomb interaction
vanishes for large momenta.  We only consider one spin species.  The
standard (Lippman-Schwinger) equation for scattering states \cite{merzbacher1970}
in momentum space takes the form
\begin{equation}
\psi^{(+)}(\vec{p})=\psi^{(0)}(\vec{p})-G^{(0)}(\vec{p})
\int\frac{d^2p'}{(2\pi)^2}V(\vec{p}-\vec{p}')\psi^{(+)}(\vec{p}'),\label{psi_k-space}
\end{equation}
where $G^{(0)}$ is the unperturbed (matrix) Green's function
and $\psi^{(0)}$ is an  eigenstate for $V=0$.
Eq.~(\ref{psi_k-space})
is conveniently recast in terms of angular components, for which we define
$
\psi_m(p)=\int^{2\pi}_0\frac{d\theta_p}{2\pi}e^{-i m \theta_p}\psi(\vec{p})$,
and decompose the Coulomb interaction in the form
$
V(|\vec{p}-\vec{p}'|)=\sum_n e^{-i n (\theta_{\vec{p}}-\theta_{\vec{p}'})}f_n(p'/p)/p,
$
where
\begin{equation}
f_m(x)=\int^{2\pi}_0\frac{d\theta}{2\pi}\frac{e^{-im\theta}}{[1+x^2-2x\cos(\theta)]^{1/2}}.
\end{equation}
In terms of these quantities one arrives at a set of equations
coupling $\psi^{(+)}_{1,m}$ and $\psi^{(+)}_{2,m+1}$, where $1,2$ are
sublattice indices. To search for power law behavior in
$\psi^{(+)}(p)$ at large momentum $p$, we adopt the
ansatz
$\psi^{(+)}_{\alpha,m}(p)={c_{\alpha,m}}/{p^s}$ for  $p\rightarrow\infty$.
To lowest order in $1/p$,
non-vanishing solutions are supported when
\begin{equation}
1-(Z\beta)^2I_m(s)I_{m+1}(s)=0,
\label{nonint-eq}\end{equation}
where $I_m(s)=\int^\infty_0x^{1-s}f_{-m}(x)dx.$ Note that $I_m(s)$
diverges for real $s$ when $s=1-|m|$ and $s=2+|m|$, and has a positive
minimum value at $s=3/2$.  From these properties, one may
see that Eq.~(\ref{nonint-eq}) has solutions with real $s$ when
$Z\beta$ is below a critical $(Z\beta)_c$; above this $s$ becomes
complex.  In particular, for $m=0$, $(Z\beta)_c=1/2$.  This is Coulomb
implosion expressed in momentum-space, which may be readily translated
into a many-body context.

{\it Power Law Behavior for Interacting Dirac Electrons} --
The same analysis for a single particle-hole pair (the
analog of the Cooper problem in superconductivity) reveals a
short-distance power law changing from real to complex at a
critical coupling \cite{wf_unpub}.  For the fully interacting
many-body problem, such behavior should be contained in the four-point
vertex function.  Since the result is non-analytic in momentum, a
nonperturbative approach is necessary.  As we now demonstrate, the
simplest relevant approximation scheme (a ladder sum \cite{fetter})
shows a change in the exponent from real to complex in the
particle-hole channel as $\beta$ is increased.  While a
generic four point vertex includes the non-analytic behavior, the
nonanalyticity cancels in the density-density response \cite{sym}.
By contrast, a potential that couples to the two sublattices
asymmetrically (for example, by a substitutional impurity, or due
to interaction with a substrate), will
include a response that is {\it antisymmetric} in the sublattice
densities.  This antisymmetric response displays the power law behavior.
For the remainder of our
discussion we focus on this antisymmetric response function.

\begin{figure}
\begin{center}
\subfigure[]{\label{fig:3leg_diagram}\includegraphics[scale=0.45]{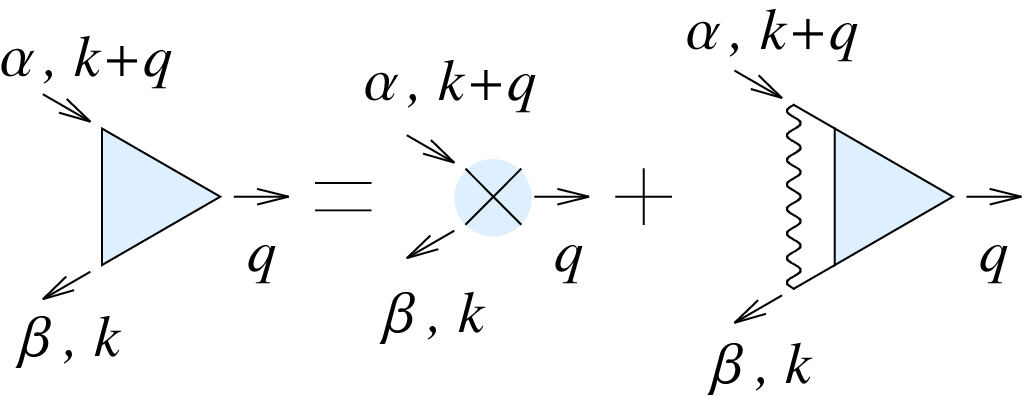}}
\hspace{5mm}
\subfigure[]{\label{fig:Mq_diagram}\includegraphics[scale=0.145]{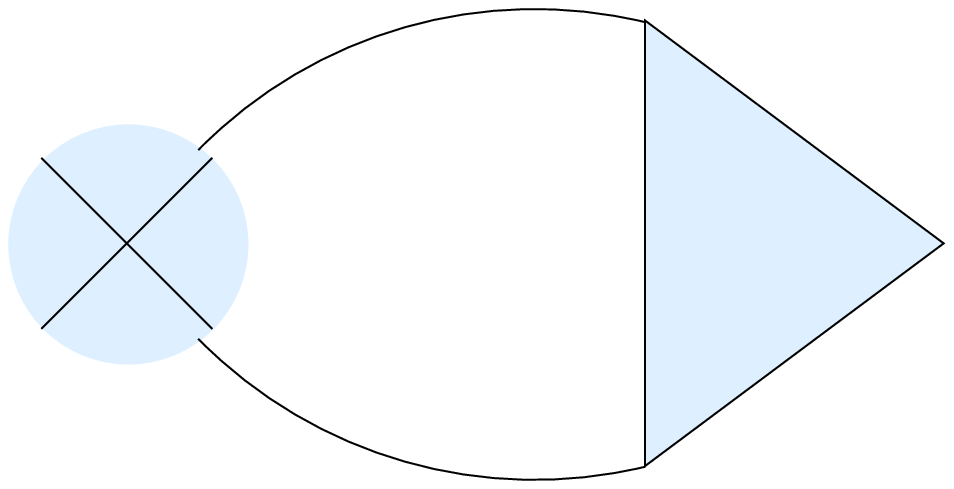}}
\end{center}
\caption{\label{FeynmanDiagrams} (Color online) (a) Diagrammatical equation for the 3-leg vertex $\tilde{\Gamma}^M_{\alpha\beta}(\vec{k},\vec{q})$ with $\sigma^z$ as the zeroth order vertex (the shaded cross in the figure). (b) Diagram for $M(\vec{q})$.
}
\end{figure}

If the sublattice index is viewed as a pseudospin, the antisymmetric
response can be viewed as a generalized magnetic susceptibility, defined as
\begin{eqnarray}
M(\vec{q})=-\frac{i}{A}\int^\infty_0dt\left<[\hat{m}_z(-\vec{q},t),\hat{m}_z(\vec{q},0)]\right>,\\
\hat{m}_z(\vec{q})=\sigma^z_{\alpha\beta}\hat{\rho}_{\alpha\beta}(\vec{q}),\mbox{       }
\hat{\rho}_{\alpha\beta}\equiv \sum_{\vec{k}}a^\dagger_{\vec{k}+\vec{q},\alpha}a_{\vec{k},\beta},
\label{M(q)}\end{eqnarray}
where $A$ is the area of the sample, repeated indices are summed,
and henceforward we will set $\hbar=1$. It can be shown that the reducible diagrams vanish in the computation of $M(\vec{q})$, so that we only need to consider irreducible bubble diagrams [Fig.~\ref{fig:Mq_diagram}].

The equation for the 3-leg antisymmetric vertex [Fig.~\ref{fig:3leg_diagram}] is
\begin{eqnarray}
\lefteqn{\tilde{\Gamma}^M_{\alpha\beta}({\vec k},{\vec q})=\sigma^z_{\alpha\beta}+}\\ \nonumber
&&\int\frac{d^2q'}{(2\pi)^2}
U(|\vec{q'}|)
\tilde{K}_{\alpha\beta\gamma\delta}(\vec{k}-\vec{q}',\vec{q})\tilde{\Gamma}^M_{\gamma\delta}(\vec{k}-\vec{q}',\vec{q}),
\end{eqnarray}
where $U(|\vec{q}|)=2\pi e^2/|\vec{q}|$,
and
\begin{equation}
\tilde{K}_{\alpha\beta,\mu\nu}(\vec{p},\vec{q})\equiv i\int\frac{dp_0}{2\pi}G^{(0)}_{\mu\alpha}(p+q)G^{(0)}_{\beta\nu}(p).
\end{equation}
Note that on the right-hand side of this expression, momenta are 3-vectors ($q_0=0$), whereas
elsewhere only the spatial components of the momenta remain.
Defining
$
\tilde{\chi}^M_{\alpha\beta}(\vec{k},\vec{q})=\tilde{K}_{\alpha\beta\gamma\delta}(\vec{k},\vec{q})\tilde{\Gamma}^M_{\gamma\delta}(\vec{k},\vec{q}),
$
one finds
\begin{eqnarray}
\lefteqn{\tilde{\chi}^M_{\alpha\beta}(\vec{k},\vec{q})=\tilde{K}_{\alpha\beta\gamma\delta}(\vec{k},\vec{q})\sigma^z_{\gamma\delta}}\\ \nonumber
&&+\tilde{K}_{\alpha\beta\gamma\delta}(\vec{k},\vec{q})\int\frac{d^2q'}{(2\pi)^2}U(|\vec{q}'|)\tilde{\chi}^M_{\gamma\delta}(\vec{k}-\vec{q}',\vec{q}).
\end{eqnarray}
This quantity is related to the susceptibility by
\begin{equation}
M(\vec{q})=\int\frac{d^2k}{(2\pi)^2}\sigma^z_{\alpha\alpha}\tilde{\chi}^M_{\alpha\alpha}(\vec{k},\vec{q}).
\end{equation}
Henceforth we focus on the long wavelength limit (small $q$), so for the moment we
drop all terms of $O(q^2)$ and higher.  Using a circular
moment expansion, one finds
\begin{eqnarray}
\lefteqn{\tilde{\chi}^{M(0)}_{\underline{\alpha\alpha}}(\vec{k},\vec{q})=\tilde{K}^{(0)}_{\underline{\alpha\alpha}\beta\beta}(\vec{k},\vec{q})\sigma^z_{\beta\beta}}\label{gen_q=0_eq}\\ \nonumber
&&+\tilde{K}^{(0)}_{\underline{\alpha\alpha}\beta\beta}(\vec{k},\vec{q})\frac{\beta}{k}\int^\Lambda_{k_0} k'dk'f_0(\frac{k'}{k}) \tilde{\chi}^{M(0)}_{\beta\beta}(\vec{k}',\vec{q}).
\end{eqnarray}
Here we used the superscript $(0)$ to denote the circular component
$m=0$, and the underlined indices are not summed over.  In this
equation we have introduced explicit ultraviolet ($\Lambda\sim
2\pi/a$, $a$ = lattice spacing) and infrared ($k_0\sim 2\pi/L$, $L$ =
linear size of system) cutoffs.

Defining $
\tilde{\chi}^{M(0)}(k,\vec{q})\equiv\sigma^z_{\beta\beta}\tilde{\chi}^{M(0)}_{\beta\beta}(k,\vec{q}),
$ in the limit $q\rightarrow0$ the solution to Eq.~(\ref{gen_q=0_eq})
may be written in the form
$\tilde{\chi}^{M(0)}(k,0)=\frac{1}{v_Fk}F(\frac{k}{\Lambda}),$
where $F$ obeys the integral equation
\begin{equation}
F(\frac{k}{\Lambda})=1+\frac{\beta}{2k}\int^\Lambda_{k_0}dk'f_0(\frac{k'}{k})F(\frac{k'}{\Lambda}).
\label{Fwithq0}
\end{equation}
Note that $F$ depends on the ratio $k/\Lambda$, a
reflection of the fact that the original Hamiltonian has no intrinsic
length scale, so (in the limit $k_0 \rightarrow 0$) $k$ can enter only
in this ratio.  For $k/\Lambda\ll 1$, one easily confirms that
Eq. (\ref{Fwithq0}) is solved by a power law $F(\frac{k}{\Lambda}) \sim (\Lambda/k)^s$,
with $s$ going from
real to complex above some critical $\beta$.

\begin{figure}
\begin{center}
\subfigure[]{\label{fig:Fkq0b03}\includegraphics[scale=0.65]{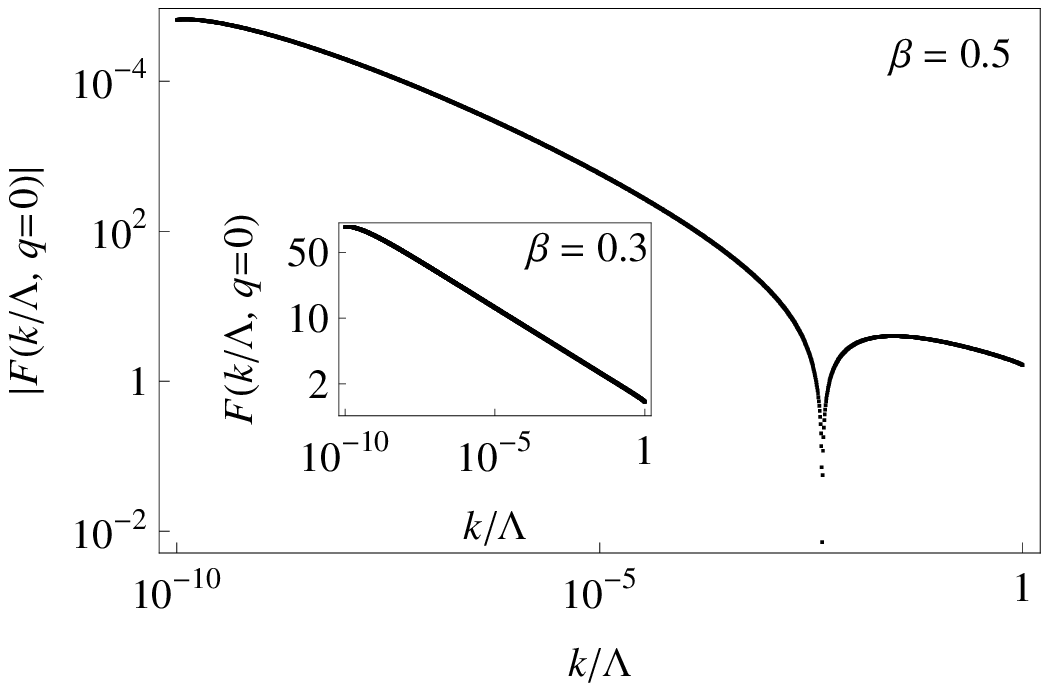}}
\subfigure[]{\label{fig:Mq0}\includegraphics[scale=0.65]{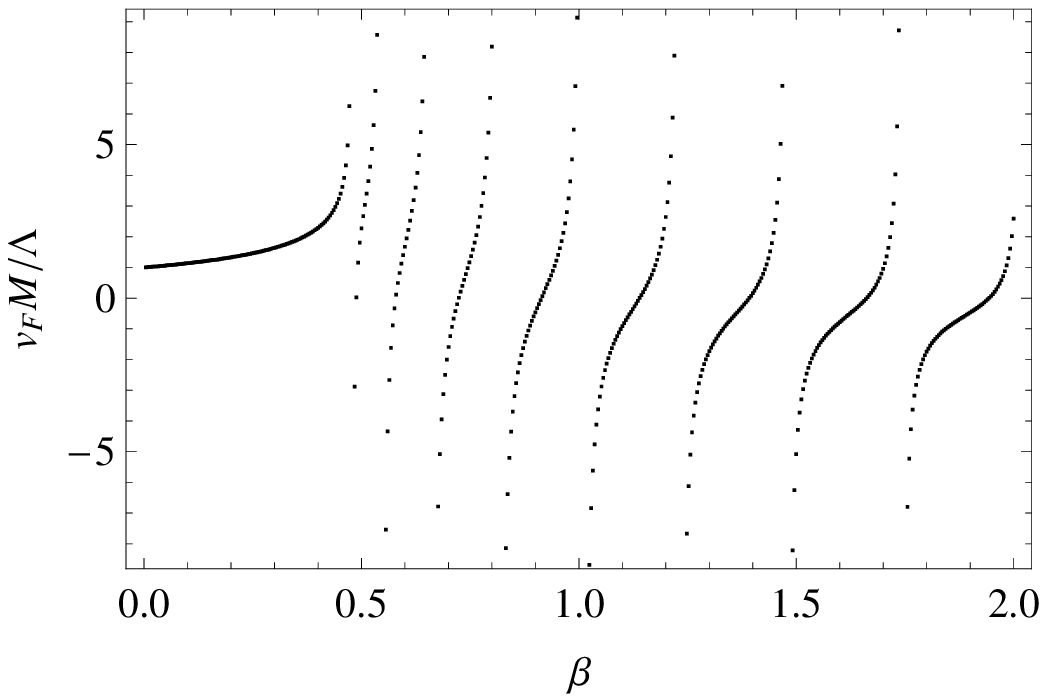}}
\end{center}
\caption{\label{q0results}Solutions of Eq.~(\ref{Fwithq0}) with $k_0/\Lambda=10^{-10}$. (a) $\beta=0.5$. Because the plotted is $|F|$, the oscillations appear as cusps. Note the amplitude of the oscillation scales roughly as $1/\sqrt{k}$. {\it Inset:} $F$ for $\beta=0.3$. It is clearly a power law except for $k$ close to $k_0$. (b) The antisymmetric response $M$ as a function of the interaction strength $\beta$.
}
\end{figure}

To confirm this, we solved Eq.~(\ref{Fwithq0}) numerically. For small
$\beta$, the solution is indeed a power law, provided $k\gg k_0$ [see
Fig.~\ref{fig:Fkq0b03} inset].  For large enough $\beta$, the
solution is consistent with a power law of complex exponent, such that
$F$ becomes oscillatory with a power law envelope
[Fig.~\ref{fig:Fkq0b03}].  Moreover, $ M(\vec{q}\rightarrow
0)=\int\frac{d^2k}{(2\pi)^2}\tilde{\chi}^M(k,\vec{q}\rightarrow 0) $ has
a series of divergences [Fig.~\ref{fig:Mq0}].
The positions and weights
of these poles depend on $k_0$.  We return to this
important point momentarily.

\begin{figure}[t]
\begin{center}
\includegraphics[scale=0.75]{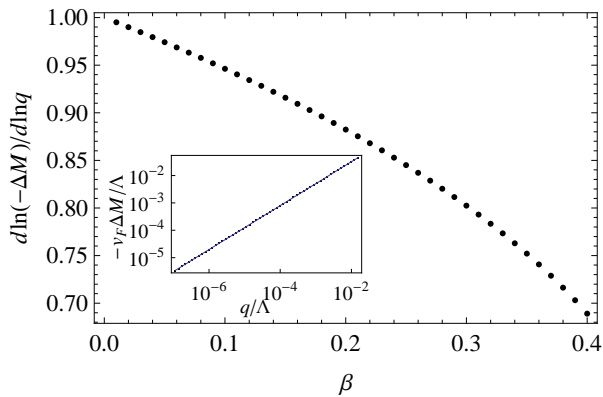}
\end{center}
\caption{\label{fig:dM}The exponent in $\Delta M(q)$ as a funciton of $\beta$. {\it Inset:} $\Delta M(q)$ for $\beta=0.3$.
}
\end{figure}

For small but nonzero $q$, it is interesting to compute the correction
$\Delta M(q)=M(q)-M(0)$.  The equation for the corresponding $\Delta
F$ has a form very similar to Eq.~(\ref{Fwithq0}), with only the
``$1$'' replaced by an inhomogeneous term, which is proportional to
$q^2/k^2$ for $k \gg q$.  The $\Delta M(q)$ resulting from this then
vanishes with an exponent that varies with $\beta$.  The inset of Fig.~\ref{fig:dM}
illustrates a typical result for $\beta$ not too large; the
exponent as a function of $\beta$ is illustrated in Fig.~\ref{fig:dM}.
One physical consequence of this is that
the difference in charge between
sublattices for an impurity placed asymmetrically with respect
to the sublattices will fall off with a $\beta$-dependent power law at
large distances, behavior which
may be observable with a local scanning probe.  We note that $\Delta M(q)$
has singularities at the same values of $\beta$ as $M(0)$, as should
be expected from the form of Eq.~(\ref{Fwithq0}).

Figs. \ref{q0results} and \ref{fig:dM} are the central results of our
work. They demonstrate that in the ladder approximation for the
many-body problem of interacting electrons in undoped graphene: (i)
For $\beta<\beta_c$ generic particle-hole correlators decay with a
power law at long distances, with an exponent varying continuously
with $\beta$.
The weak-coupling many-body groundstate thus displays a basic
property of a critical phase.
The power law behavior is directly visible
in the sublattice-antisymmetric density correlator.
(ii) For $\beta>\beta_c$ the exponent becomes complex, as in the
noninteracting Coulomb implosion problem. In the interacting many-body
case, the susceptibility $M(q)$ of
Eq. (\ref{M(q)}) diverges for $k_0 >
0$. This strongly suggests a quantum phase transition
to broken symmetry state with
staggered charge
order \cite{khveshchenko,gorbar2002}. 
However, the presence of {\it many} such divergences
as a function of $\beta$ suggests there are different
ways to break the symmetry. 
(iii) The singularities vanish
in the thermodynamic limit, with the poles merging into
a continuous function.  The separation between
them vanishes only
logarithmically as
$k_0 \rightarrow 0$, as we demonstrate below, resulting in a
branch point at $\beta_c$.  We interpret this latter non-analytic behavior
as the signal of a phase transition.
Since it is a result of the merging poles, a natural interpretation is
that the instability is into a state involving fluctuations among 
different realizations of a chiral order parameter which, if static, would
produce a gapped exciton phase \cite{khveshchenko,gorbar2002}.
We speculate that with further increase in $\beta$, one of these orderings
could be favored over the others, resulting in a true condensed phase.


{\it Analytical Results for Model Kernel} --
A fuller understanding of Eq.~(\ref{Fwithq0}) may be
obtained
by adopting a model
kernel,
\begin{equation}
\tilde{f}_0(x)=\theta(1-x)+\frac{1}{x}\theta(x-1).
\end{equation}
This has the same behavior as the real kernel at large and small $x$,
and is simple enough to allow analytic solutions.
We have verified numerically that the results for $F$ and $M$ are
qualitatively very similar to those obtained with the correct $f_0$.
With this kernel, Eq.~(\ref{Fwithq0}) has general solutions of the form
\begin{equation}
F(\tilde{k})=A_+ \tilde{k}^{\lambda_+}+A_- \tilde{k}^{\lambda_-},
\label{model_F}
\end{equation}
with $\tilde{k}=k/\Lambda$,
$\lambda_{\pm}=\frac{-1\pm\gamma}{2}$, and $\gamma=\sqrt{1-2\beta}$.
The coefficients $A_{\pm}$ are determined by substituting Eq.~(\ref{model_F})
back into the integral equation.  This results in power law behavior
for $k \gg k_0$, with exponent $\lambda_+$, which goes from real to
complex when $\beta$ exceeds 1/2.  Moreover, $M(q \rightarrow 0)$ may
be evaluated, yielding
\begin{equation}
M(0)=\frac{\Lambda}{v_F}\frac{2-2\tilde{k}_0^\gamma}{1+\gamma-\beta+\tilde{k}_0^\gamma (-1+\gamma+\beta)}.
\end{equation}
This
has poles for $\beta>1/2$ when
\begin{equation}
\sqrt{2\beta-1}\ln{\tilde{k}_0}=2\arctan{\frac{\sqrt{2\beta-1}}{1-\beta}}+2\pi n,
\end{equation}
with integer $n$ and $0<\arctan{(x)}<\pi$.  Note that the
distance between poles vanishes logarithmically as $\tilde{k}_0 \rightarrow
0$, as discussed above.
{Furthermore, for $\beta>1/2$, $\tilde{k}_0^{\gamma}$
becomes ill-defined unless an infinitesimal imaginary part is introduced
in $\beta$, so that $\beta=1/2$ becomes a branch point for $M(0)$.  We interpret
this as the signal of a phase transition in the thermodynamic limit, since
$M(0)$ need not be real and positive beyond this point.}

In this work we have examined the problem of interacting electrons in
undoped graphene in the ladder approximation. For
$\beta<\beta_c$ the ground state has $\beta$-dependent power law correlations
in the antisymmetric-sublattice density response.
We note that in light of the
the logarithmic growth of $v_F$ \cite{gonzalez94,herbut08},
one may expect logarithmic corrections to this
power, which may be difficult to detect for finite size systems.
For $\beta>\beta_c$ there is a nonstandard phase
transition in which susceptibilities become divergent only for a finite size
system. The continuously varying exponent (in the ladder
approximation) and the instability
when it reaches a critical value are
reminiscent of the behavior of two-dimensional classical $XY$ models.
A central feature of our analysis is
that, due to the absence of an intrinsic length scale, short distance
power laws from anomalous penetration of a centrifugal
barrier by the Coulomb interaction have an impact
on the
long-distance decay of correlators.  Corresponding behavior
occurs for non-interacting Dirac electrons near a Coulomb
impurity.  This analogy has been
noted recently in a different way\cite{gamayun-2009}.

\begin{acknowledgments}
The authors thank L. Brey for many helpful discussions.  Support was
provided by the NSF under Grant Nos. DMR-0704033 (JW and HAF) and
DMR-0703992 (GM).  Numerical calculations described here were
performed on Indiana University's computer cluster Quarry.  
\end{acknowledgments}


\end{document}